\documentclass[conference]{IEEEtran}
\IEEEoverridecommandlockouts
\usepackage{cite}
\usepackage{amsmath,amssymb,amsfonts}
\usepackage{bbm}
\usepackage{graphicx}
\usepackage{textcomp}
\usepackage{xcolor}
\usepackage{kotex}
\usepackage{algorithmicx,algpseudocode}
\usepackage{subfigure}
\usepackage{booktabs}
\usepackage[linesnumbered,ruled]{algorithm2e}
\newcommand{\BfPara}[1]{{\noindent\bf#1.}\xspace}

\def\BibTeX{{\rm B\kern-.05em{\sc i\kern-.025em b}\kern-.08em
    T\kern-.1667em\lower.7ex\hbox{E}\kern-.125emX}}
\begin{document}

\title{Multi-Agent Deep Reinforcement Learning for Efficient Passenger Delivery in Urban Air Mobility}

\author{
\IEEEauthorblockN{Chanyoung Park$^\dag$, Soohyun Park$^\dag$, Gyu Seon Kim$^{\dag}$, Soyi Jung$^\S$, Jae-Hyun Kim$^\S$, and Joongheon Kim$^\dag$}
\IEEEauthorblockA{
$^\dag$\textit{Department of Electrical and Computer Engineering, Korea University, Seoul 02841, Republic of Korea}\\
%$^\ddag$\textit{Department of Aerospace Engineering, Inha University, Incheon 21999, Republic of Korea}\\
$^\S$\textit{Department of Electrical and Computer Engineering, Ajou University, Suwon 16499, Republic of Korea}
}
E-mails: \texttt{\{cosdeneb,soohyun828,kingdom0545\}@korea.ac.kr}, 
\texttt{\{sjung,jkim\}@ajou.ac.kr}, \\
\texttt{joongheon@korea.ac.kr}
}

\maketitle

\begin{abstract}
It has been considered that \textit{urban air mobility (UAM)}, also known as drone-taxi or electrical vertical takeoff and landing (eVTOL), will play a key role in future transportation. By putting UAM into practical future transportation, several benefits can be realized, i.e., (i) the total travel time of passengers can be reduced compared to traditional transportation and (ii) there is no environmental pollution and no special labor costs to operate the system because electric batteries will be used in UAM system. However, there are various dynamic and uncertain factors in the flight environment, i.e., passenger sudden service requests, battery discharge, and collision among UAMs. Therefore, this paper proposes a novel cooperative multi-agent deep reinforcement learning (MADRL) algorithm based on centralized training and distributed execution (CTDE) concepts for reliable and efficient passenger delivery in UAM networks. According to the performance evaluation results, we confirm that the proposed algorithm outperforms other existing algorithms in terms of the number of serviced passengers increase (30\%) and the waiting time per serviced passenger decrease (26\%).
\end{abstract}


\section{Introduction}
% UAM을 이용한 drone taxi or air taxi가 등장. 예시) Uber Elevate (acquired by Joby Aviation), Amazon Prime Air, NASA's Unmanned Traffic Magement (UTM). 공중에서 vehicle간 경로 방해나 충돌을 방지하기 위해 UAM의 이동과 제어, 조작에 대한 연구가 많이 진행되고 있음. 특히 UAM 기반의 taxi 시스템의 경우 승객을 출발지점에서부터 목적지가지 이동시킴에 있어 최적 서비스를 제공하기 위한 기술이 필요함. 
Many studies have been conducted on urban air mobility (UAM) such as Uber Elevate (acquired by Joby Aviation)~\cite{jobyaviation}, Amazon Prime Air~\cite{Amazon}, and NASA's unmanned traffic management (UTM)~\cite{UTM}. 
In order to activate UAM-based air transportation services, it is vital to conduct research on autonomous UAM mobility control and operation for preventing path/trajectory planning/interference or collision between vehicles in the air~\cite{9620751,icc,access202106park,jcn2022lee}.
In particular, in the case of a UAM-based taxi system, advanced intelligent technologies are required in terms of the number passenger delivery increase and the waiting time per serviced passenger decrease. 

%우리는 다수의 UAM과 승객, 일정 간격으로 배치된 vertiport가 있는 Air Transportation service 환경을 가정하고, UAM 기반의 drone taxi 서비스를 제공하는 도심 환경에서 다수의 UAM이 서로 협력적으로 시스템 내에 존재하는 탑승객에게 최적의 서비스를 제공하는 것을 목적으로 함. 이 때, 사용자들의 위치와 원하는 목적지가 모두 상이하고, UAM는 사용자들이 현재 탑승을 대기하고 있는 vertiport에서 target vertiport로 user를 이송한다. UAM은 vertiport를 이동하며 서비스를 제공하는데, 승객이 탑승하거나 하차하기 위해서 vertiport에 머무는 동안 충분한 양의 power를 충전받을 수 있는 환경으로 가정한다. 
This paper assumes an environment with multiple UAM, passengers, and vertiports deployed at regular intervals~\cite{9620751}. We aim to provide optimal service to all passengers through new UAM-based air transportation services. The passengers' locations and desired destinations are different at this time. The UAM transfers the passenger by moving the vertiport where the passenger is waiting to board to the target vertiport. In addition, it is assumed that sufficient power can be charged while the UAM stays at the vertiport for passenger boarding or getting off. 
%본 논문에서는 앞서 언급한 Multi-UAM 기반의 air transportation service를 위해 심층 강화학습 기법을 적용하고자 함. UAV를 활용한 네트워크에서 강화학습을 활용한 control or trajectory optimization 연구들은 강화학습이 air mobility control에 적합한 기술이라는 것을 다양한 결과와 성능 개선 효과로 보여주고 있음.  (i)UAM의 power 상태, (ii) air transportation service 를 받은 user의 수, and (iii) user들의 대기 시간을 보상함수로 고려.
To achieve the purpose of multi-UAM based high level air transportation service, we proposed a new UAM and passenger matching algorithm using deep reinforcement learning (DRL). 
There are many studies using DRL for optimizing control or trajectory in a network using unmanned aerial vehicles (UAVs), and these studies show results and performance improvement effects that reinforcement learning is a suitable technology for air mobility control.
For this reason, Three factors are considered as components of reward function in DRL-based proposed algorithm, e.g., (i) the power state of the UAM, (ii) the number of users who received the air transportation service, and (iii) the latency of users.
% 하나의 agent에서 동작되는 일반적인 심층 강화학습 알고리즘 (e.g., DQN, DDPG, etc.)를 사용하는 경우 UAM 하나가 관찰할 수 있는 환경 정보가 매우 제한적이기 때문에 '시스템 전체 (이 표현을 바꾸고 싶음)'에 대한 고려 없어 UAM간 cooperation이  불가능하다. 이러한 이유로 CommNet 을 적용함으로써 앞서 언급한 세 가지 보상함수 요소들 중 proposed algorithm을 통해 모든 UAM 협력적으로 달성해야 하는 (iii) 짧은 user 대기 시간을 common reward로, 개별 UAM에서의 energy와 UAM이 서비스한 user수를 individual reward로 구분한다.  결과적으로 시스템에 존재하는 UAM간 information sharing을 통해 협력적 air-transportation service 제공이 가능하다. 
However, in the case of using general DRL algorithms that operate only with environmental information observed by one agent, cooperation between several UAMs is impossible due to the lack of consideration for the system-wide efficiency. For this reason, we design our proposed algorithm inspired by multi-agent deep reinforcement learning (MADRL), i.e., \textit{CommNet}~\cite{tvt202106jung, tii202210yun, park2022cooperative,tii2105commnet}, which can share observation information for inter-agent communications. We also divide the reward components as a common reward (e.g., (iii) short user latency) to be achieved collaboratively in all UAMs and individual reward (e.g., (i) the power state and (ii) the number of users serviced) got independently from one UAM. As a result, the cooperative air transportation service can be provided through information sharing between UAMs.

%MADRL/CommNet 기반의 UAM 간 통신을 이용한 협업적인 air transportation 서비스를 제안한다.
%centralized training and distributed execution (CTDE) for multi-agent cooperation.
%실제 UAM 모델의 사양의 고려하고, 자세한 reward 함수 설계를 통해 실제 환경과 유사한 MADRL 환경을 구축하였다.
%다양한 성능 평가를 통해 제안하는 알고리즘이 가장 좋은 성능을 내는 것을 증명하였다.

\begin{figure}
    \centering
    \includegraphics[width=0.99\linewidth]{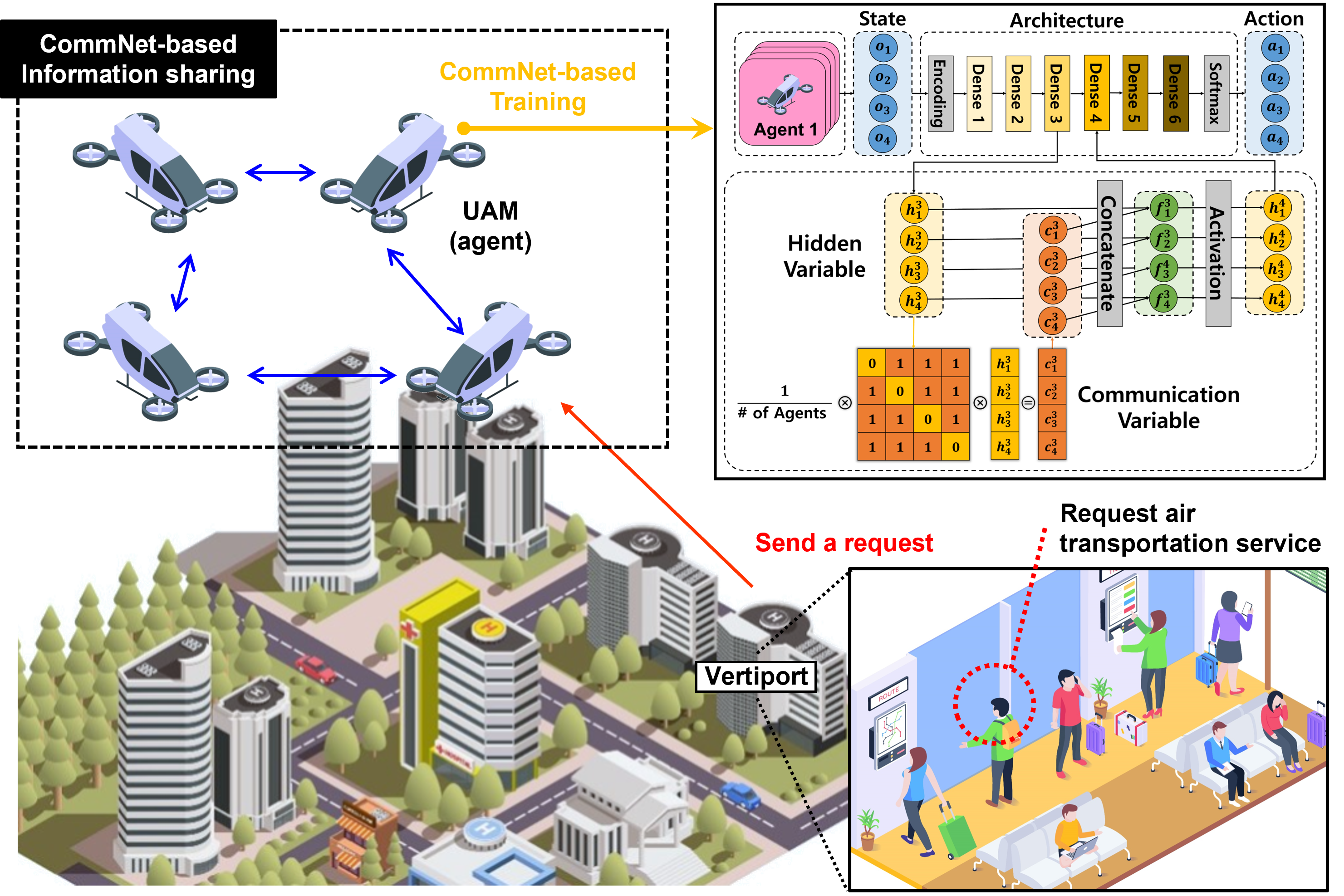}
    \caption{Overall system model of UAM-based air transportation service.}
    \label{fig:UAM Scenario}
    \vspace{-5mm}
\end{figure}

The main contributions are as follows, i.e., (i) we propose a CommNet-based cooperative air transportation service algorithm via communication between multi-UAMs. The proposed algorithm is suitable for the environment which needs multi-agent cooperation when a single agent's observation range is limited compared to system-wide information using CTDE; and (ii) to analyze the performance of the proposed algorithm, we consider the specifications of the real-world UAM model, build an MADRL environment similar to the real-world environment, and design a reward function suitable for the system. %Furthermore, we prove that the proposed algorithm performs the best through various performance evaluations.

% \section{Preliminaries}
% \subsection{Reference Network Model}
% https://www.overleaf.com/project/63258bec5d30b5f8efabfbd8
% \subsection{Related and Previous Work}

\section{MADRL for Efficient Passenger Delivery in UAM Networks}
\subsection{Algorithm Design}
\subsubsection{Reinforcement Learning Formulation}
We formulate our MADRL problem with the Markov decision process (MDP). MDP follows the Markov property that the present determines the future, regardless of the past. The components of MDP are $\left(\mathcal{S}, \mathcal{A}, \mathcal{R}, \mathcal{P}, \gamma\right)$, which are state space, action space, reward function, transition probability, and discount factor, respectively. The state space $\mathcal{S}$ and action space $\mathcal{A}$ are sets of states and actions that an agent can have, and the reward function $\mathcal{R}$ is a function that outputs a reward when an agent in a specific state takes a specific action. The goal of reinforcement learning is to find a policy matching the state with action which maximizes reward in an episode. The transition probability $\mathcal{P}$ is the probability that an agent will reach the next state from its current state. The discount factor $\gamma$ has a value between 0 and 1 and determines how much the agent weights the rewards it will get in the future compared to now. Here, $\gamma$ helps to train optimal policy faster. In the followings, we will describe how we designed the MDP in this paper.

\BfPara{State}
In this paper, we denote the sets of passengers, UAM agents, and vertiports as
$\mathcal{U} \triangleq \left\{ u_{1}, \dots, u_n, \dots, u_{N} \right\}$, 
$\mathcal{B} \triangleq \left\{ b_{1}, \dots, b_m, \dots, b_{M} \right\}$,
$\mathcal{V} \triangleq \left\{ v_{1}, \dots, v_k, \dots, v_{K} \right\}$
where $u_n$ is $n$-th passenger, $b_m$ is $m$-th UAM agent, and $v_k$ is $k$-th vertiport, respectively. The state space corresponds to the environmental information observed by the UAM agent. The state space that $b_m$ can observe is comprised of three parts, which are denoted to $S_m\in\{\mathcal{S}^l_m, \mathcal{S}^s_m, \mathcal{S}^e_m\}$, where $\mathcal{S}^l_m$ is the location information, $\mathcal{S}^s_m$ is the air transportation service information, and $\mathcal{S}^e_m$ is the energy information related to UAM agent's battery state. Therefore, The set of all UAM agents' state space can be denoted as $\mathcal{S}\triangleq\{\mathcal{S}_1,\cdots,\mathcal{S}_m,\cdots\mathcal{S}_M\}$. The details of above three parts are as follows. Note that all UAM agents can observe partial information around their locations due to the physical limitations.
\begin{itemize}
\item \textit{Location Information:}
We denote the position of $b_m$ as $p(b_m)\triangleq\{x_m, y_m, z_m\}$ in 3D Cartesian coordination system. The UAM agent also has to consider the distance from other UAM agents to avoid collisions among UAM agents. We denote the distance with $m$-th UAM agent and other UAM agent as $d(b_m, b_{m'})$, where $b_m$ and $b_{m'}$ are not same UAM agents. In addition, the UAM agent has to know the location of vertiports to serve air transportation service. The distance of $b_m$ and $v_k$ are denoted as $d(b_m, v_k)$. In a nutshell, the location information of $b_m$ is $\mathcal{S}^l_m\triangleq\{p(b_m), d(b_m, b_{m'}), d(b_m, v_k)\}$.
\item \textit{Service Information:}
All UAM agents must consider information about their services to provide reliable air transportation service to passengers. In our proposed model, $u_n$ randomly sends request $\chi_n$ for passenger transportation service at a specific time step $t$ in any $v_k$. Suppose $b_m$ is within the cell radius of $v_k$ where $u_n$ who sends a request is located. In that case, $b_m$ can accept the request of $u_n$ and establish a wireless link. 
Therefore, $\chi_{mn} = 1$ when $d(b_m,u_n) \leq \zeta(v_k)$ 
%\begin{equation}
%\chi_{mn} =
%\begin{cases}
%    1, & \mbox{if  } d(b_m,u_n) \leq \zeta(v_k),\\
%    0, & \mbox{otherwise},
%\end{cases}
%\end{equation}
where $\zeta(\cdot)$ is the function which outputs the cell radius of $v_k$ (otherwise, $\chi_{mn} = 0$). We also denote the status of the link, seat, and waiting time for passengers on board of $b_m$ by $\eta_m\triangleq\{\lambda^{1}_{n},\cdots,\lambda^{\varsigma}_{n'}\}$, $\Psi_m\triangleq\{\psi^{1}_{n},\cdots,\psi^{\varsigma}_{n'}\}$, and $\tau_m\triangleq\{\omega^{1}_{n},\cdots,\omega^{\varsigma}_{n'}\}$,
$\forall \{n, n'\} \in [1, N]$, respectively. $\varsigma$ is the size of the UAM agent's seat, $\lambda^1_n$ is the established wireless link between $u_n$ and $b_m$, and $u^{1}_n$ means $u_n$ is boarding on the first seat. In addition, $\lambda_{n}$, $\lambda_{n'}$ are for distinguish between other links, and $\psi_{n}$, $\psi_{n'}$ are for distinguish between $n$-th passenger and other passengers. Once the connected passenger boards the UAM agent, the established link is maintained until they reach their destination vertiport. However, the established radio link is broken if the UAM agent leaves the vertiport's cell radius without picking up the connected passenger. The waiting time for passengers is the time from starting to send a request to arrive at the destination vertiport. All UAM agents try to reduce the overall passengers' waiting time to provide reliable service. To sum up, the service information of $b_m$ is $\mathcal{S}^s_m\triangleq\{\cup_n^N\{\chi_{mn}\}, \eta_m, \Psi_m, \tau_m\}$.
\item \textit{Energy Information:}
Each UAM agent needs to consider its energy information to prevent full discharging for reliable service. At every time step $t$, UAM agents consume operational energy $e_{m}$ that is the amount of energy remaining of $b_{m}$, which is a component of energy information $\mathcal{S}^e_m\triangleq\{e_m\}$. We assume that all UAM agents are fully charged. UAM agent is an electrified air vehicle that uses a battery unlike an internal combustion engine-based aircraft that uses fossil fuels. Thus, $e_{m}$ can be expressed through the aerodynamically calculated power without considering the specific fuel consumption. Here, ~\eqref{eq:hovering energy} is the power when the UAM agent is ascending or descending while carrying passengers~\cite{tvt202106jung}, i.e.,
\begin{equation}
P_h=\underbrace{\frac{\delta }{8}\rho sA\Omega^3R^3}_{\textrm{bladeprofile}, \,P_{o}}+\underbrace{(1+k)\frac{W^{\frac{3}{2}}}{\sqrt{2\rho A}}}_{\textrm{induced}, \,P_{i}},
\label{eq:hovering energy}
\end{equation}
where $\delta$, $\rho$, $s$, $A$, $\Omega$, $k$, $R$, and $W$ are the drag coefficient, air density, rotor solidity, rotor disc area, blade angular velocity, induced drag coefficient, rotor radius, and total weight considering the passenger's payload, respectively. In~\eqref{eq:hovering energy}, the blade profile $P_{0}$ is the power required to rotate to the blade, and the induced $P_{i}$ is the power to overcome the induced drag which is caused by lift due to the finite wingspan. Next, the energy expenditure for round-trip traveling of $b_m$ is expressed as follows~\cite{zeng2017energy},

\begin{table}[t]
\scriptsize
\centering
\caption{Specification of UAM Model.}
\renewcommand{\arraystretch}{1.0}
\begin{tabular}{c||c}
\toprule[1pt]
\textbf{Notation} & \textbf{Value} \\ \midrule
Maximum number of passengers, $\varsigma$ & 4 \\
Flight speed, $v$ & 73.762 m/s \\
Aircraft mass including battery and propellers, $m$ & 1815 kg\\
Aircraft weight including battery and propellers, $W=mg$ & 17799 N\\ Rotor radius, $R$ & 1.45 m \\
Rotor disc area, $A=\pi R^{2}$ & 6.61 $m^{2}$ \\
Number of blades , $b$ & 5 \\
Rotor solidity, $s=\frac{0.2231b}{\pi R}$ & 0.2449 \\
Blade angular velocity, $\Omega$ & 78 radius/s\\
Tip speed of the rotor blade , $U_{tip}=\Omega R^{2}$ & 112.776 m/s \\
Air density, $\rho$ & 1.225 kg/$m^{3}$ \\
Fuselage drag ratio, $d_{0}=\frac{0.0151}{sA}$ & 0.01 \\
Mean rotor-induced velocity in hovering, $v_{0}=\sqrt\frac{W}{s\rho A}$ & 26.45 m/s \\
Profile drag coefficient, $\delta$ & 0.045 \\
Incremental correction factor to induced power, $k$ & 0.052 \\
\bottomrule[1pt]
\end{tabular}
\label{tab:parameters of uam}
\end{table}

% The blade profile $P_0$ is the power required to rotate the blade, and the induced is the power to overcome the induced drag. Induced drag is caused by lift because it is a 3D wing with a finite wingspan. It is the drag force created by the horizontal component force generated as the lift is pushed back by the downwash-induced drag by the wing tip vortex generated at the tip of the wing. Wing tip vortex is turbulence that occurs when high-pressure air from the underside of the wing rises to the top of the wing with low pressure, which is an essential property for the characteristics of a flying vehicle. 
 
% $\delta$ is drag coefficient, which is a dimensionless coefficient that quantifies the drag force of a body in flow, $\rho$ is the density of air that decreases exponentially with altitude, and $s$ is rotor solidity, the ratio of rotor blades area to rotor disc area. $A$, $\Omega$, $R$, $W$ are rotor disc area, blade angular velocity, rotor radius and total weight considering the passenger's payload, respectively. $k$ is an induced drag coefficient inversely proportional to the efficiency factor(e) and aspect ratio (AR). To overcome the drag increased by the induced drag, the UAM must rotate the rotor, so $k$ must be considered in the power equation. When a UAM carrying a passenger rises to an altitude of 600m by eVTOL and propels forward, the force on the x-axis is added, so the propulsion power consumption $P_p(v)$ must be considered. $P_p(v)$ is expressed as

\begin{multline}
P_p= \underbrace{P_i\left(\sqrt{1+\frac{v^4}{4v_0^4}}-\frac{v^2}{2v_0^2}\right)^{0.5}}_{\textrm{induced}}\\+ \underbrace{P_0\left(1+\frac{3v^2}{U_{tip}^2}\right)}_{\textrm{bladeprofile}}+\underbrace{\frac{1}{2}d_0\rho sAv^3}_{\textrm{parasite}},
\label{eq:round trip traveling energy}
\end{multline}
where $v$, $v_0$, $U_{tip}$, and $d_0$ are UAM agent's cruising flying speed, mean rotor-induced velocity which is the average velocity of the flow driven by the wing tip vortex, tip speed of the rotor blade, and fuselage drag ratio, respectively. Aerodynamic parameters are for Joby aviation S4 model, and their values are shown in Table~\ref{tab:parameters of uam}.

According to Joby aviation's analyst day presentation~\cite{jobyaviation}, it takes about 30 seconds for S4 to ascend to its target altitude of 600m after carrying passengers. This is a value predicted through the rate of climb calculated using the wind tunnel test results of an aircraft with similar specifications to S4. At an altitude of 600m, the air density $\rho$ is similar to that at sea level, and the rate of climb of an aircraft at sea level is about 4000 ft/min. The turnaround time is up to 6 minutes, and the charge per journey is 30kWh. Upon landing, the UAM must unload passengers at each arrival vertiport within 6 minutes and rapidly charge as much as possible. The real charging time is only 5 minutes, excluding the time for passengers to disembark. S4 has four 150kWh batteries, two in the main wing inboard and two in the nacelle at the rear of the main wing inboard electric motor. The more detailed specification of the battery is shown in Table~\ref{tab:specification of UAM battery}.
 
%  To charge each battery by 30kWh in 5 minutes, the charger's power supply must be at least 360kW. Four chargers with a charging speed of 360kW supply power to each battery. When a 150kWh battery is charged with a 360kW charger, the C-rate becomes 2.4/h, and the state of charge (SOC), which is 100\%\ charging time, takes 25minutes. Since the UAM can charge 20\%\ of the total battery capacity per journey, the UAM can fly while saving enough energy. The calculation of UAM's battery specifications are shown in Table \ref{tab:specification of UAM battery}.

\begin{table}[t!]
\scriptsize
\centering
\caption{Specification of UAM Battery.}
\renewcommand{\arraystretch}{1.0}
\begin{tabular}{c||c}
\toprule[1pt]
\textbf{Notation} & \textbf{Value} \\ \midrule
Battery capacity & 150~kWh \\
Battery charge capacity per journey & 30~kWh \\
Charging time per journey & 5~min \\
Charger supply power & 360~kW \\
C-rate & 2.4 per hour \\
State of charge & 25~min \\
Charge rate per journey & 20~\%\ \\

\bottomrule[1pt]
\end{tabular}
\label{tab:specification of UAM battery}
\end{table}
\end{itemize}

\BfPara{Action}
The discrete set of actions $\mathcal{A}$ which is the UAM agent can take at each time step is composed of two types: (i) passenger transport action $\mathcal{A}_p$ (ii) takeoff/landing action $\mathcal{A}_h$. Each UAM can take action from the set of actions, i.e., $\mathcal{A} \triangleq \{ \mathcal{A}_p, \mathcal{A}_h\}$ where $\mathcal{A}_p \triangleq \{x_{pos} \pm \left(v \times t\right), y_{pos} \pm \left(v \times t\right)\}$ and $\mathcal{A}_h \triangleq \{\texttt{hovering},\; \texttt{takeoff},\;\texttt{landing}.\}$.
By taking one action, the UAM agent transitions to the next time step $t+1$. The passenger transport action $\mathcal{A}_p$ means the UAM agent moves in the air in $x$ or $y$ direction. For every time step, the UAM agent can move $\left(v \times t\right)$ in $x$ or $y$ direction, where $v$ and $t$ denote the speed of the UAM in table. \ref{tab:parameters of uam}, and current time step $t$. Next, the takeoff and landing action $\mathcal{A}_h$ means the UAM agent lands at a vertiport for boarding passengers or leaves the ground to transport passengers to their arrivals. The UAM agent can take $\mathcal{A}_h$ when there is one more link with passengers or there is a vertiport which is the arrival of one more onboard passenger. In addition, if the remaining energy of the UAM agent is not enough to cruise, the UAM agent can take $\mathcal{A}_h$ to recharge its battery. The air transportation service proceeds in the following steps: (1) Passengers send a request to receive air transportation service to UAM agents. (2) If the UAM agent is within the radius of the vertiport's communication cell, the UAM agent can accept the passenger's request and establish a wireless link with it. The established radio link is broken if the agent accepts the passenger's request but leaves the vertiport's cell radius without picking up the passenger. (3) After the agent has accepted the passenger's request and has arrived at the appropriate port, the agent can board the passenger. (4) When the agent takes off again and lands at the passenger's arrival vertiport, the passenger will disembark at that vertiport, and the air transport service will be completed.

\BfPara{Reward}
The reward is comprised of two elements: (i) individual reward (ii) common reward. The individual reward is for providing a transportation service, and the common reward is for overall transportation service quality. In every time step, agents try to maximize their total reward in~\eqref{eq:total reward}. This paper considers higher transportation service quality as less overall waiting time and more support rate.
\begin{eqnarray}
    & & 
    R^{m}(t)=\rho_s\times\underbrace{R_{c}(t)}_{\textrm{common reward}}\times\underbrace{(1+R_{i}^{m}(t))}_{\textrm{individual reward}},\label{eq:total reward}\\
    & &
    R_c(t)=\sum_{j=1}^{N}\frac{1}{1+\sum_{0}^{t}W^{j}(t)},\label{eq:common reward}\\
    & &
    R_i^m(t)= R_s^m(t)\times R_e^m(t),\label{eq:individual reward}\\
    & &
    R_e^m(t)= e_m(t) - (P_h\times\mathcal{A}_h + P_p\times\mathcal{A}_p),\label{eq:energy reward}
\end{eqnarray}
where $\rho_s$, $R_{c}(t)$, $R_{i}^{m}(t)$ are the scaling factor, common reward for all UAM agents and individual reward of $b_m$ at time step $t$, respectively. $R_{c}(t)$ is composed of the summation of all users' waiting time, where $W^{j}(t)$ is the waiting time of the $j$-th passengers from starting to send request until arriving at destination vertiport. At the end of the episode, the lower summation of all users' waiting time is, the higher the common reward UAM agents get. Consequently, all UAM agents try to reduce users' waiting time to enhance overall service quality. Next, the individual reward of $b_m$ is comprised of $R^m_s(t)$ and $R^m_e(t)$, which are the service reward and energy reward at time step $t$, respectively. Suppose $b_m$ picks up or disembarks $u_n$ (i.e., provides transportation service) in any vertiport at time step $t$. In that case, it instantaneously receives the individual reward $R^m_s(t)$ at that time step. Otherwise, $b_m$ cannot get any individual reward at that time. $R^m_e(t)$, the energy reward, is for managing its remaining battery. It is the difference between the energy information at $t$, $e_m(t)$, and energy consumption depending on the action taken by $b_m$ at time step $t$. Note that the reward components related to our environment are not considered for reward formulation because the aerial situations are generally free spaces, i.e., rare obstacles. 

\subsection{Pseudo-Code and CTDE Computation}
To cooperate with UAMs during MADRL training, we use CommNet architecture that enables inter-agent communication. Various studies are using agent-to-agent communication by CommNet for agent cooperation in MADRL~\cite{tvt202106jung, tii202210yun, park2022cooperative}. Fig.~\ref{fig:UAM Scenario} illustrates our overall air transportation systems where all UAM agents have a CommNet-based policy. Each CommNet-based UAM agent collects observation information $s$ and simultaneously shares it with other UAM agents~\cite{sukhbaatar2016learning}. Here, we set the input state as a set of UAM agent's observation information in matrix form, i.e., $s \equiv [o_1,\cdots,o_m,\cdots,o_{M}]$, which is the input of CommNet. The sequential process of CommNet architecture is expressed as, 
\begin{equation}
    h^{1}_{m} = \textsf{Encoder}(o_m),
    \label{eq:encoding}\\
\end{equation}
where $\textsf{Encoder}(\cdot)$ is an encoding function that converts observation $o_m$ to the hidden variable $h^1_m$ in the first layer, i.e.,
\begin{equation}
    c^{l}_{m} = \frac{1}{M-1} \sum_{m \neq M}\nolimits h^{l}_{m},
    \label{eq:comm}
\end{equation}

By averaging all hidden variables of all UAM agents except $b_m$ for every $l$-th layer, we can make the communication variable $c^l_m$. Note that the communication between UAM agents occurs by averaging hidden variables, i.e.,

\begin{equation}
    h^{l+1}_{m} = \textsf{Activ}(f^{l}_{m}(\textsf{Concat}(h^{l}_{m}, c^{l}_{m}))),
    \label{eq:hidden}\\
\end{equation}
where $\textsf{Activ}(\cdot)$, $f^{l}_{M}(\cdot)$, and $\textsf{Concat}(\cdot)$ are an activation function (e.g., ReLU, sigmoid, hyperbolic tangent), non-linear function of $l$-th layer, and concatenate function, respectively. The variables $\{h^l_m,c^l_m\}$ are feed-forwarded to the next layer. In CommNet, \eqref{eq:comm} and \eqref{eq:hidden} are repeatedly performed from the first layer to the last $L$-th layer, i.e.,

\begin{equation}
    {a}_m = \textsf{Softmax}(\textsf{Concat}([h^{L}_1,\cdots,h^{L}_M])),
    \label{eq:hidden2}
\end{equation}
where $\textsf{Softmax}(\cdot)$ is a softmax function. The action probability of $b_m$, which is the output of policy $\pi_m(a|s)$, is calculated by concatenating all hidden variables in the last layer $[h^{L}_1,\cdots,h^{L}_M]$ and taking a softmax function.

\begin{algorithm}[t]
%\small
%\footnotesize
%\scriptsize
    Initialize the \textit{critic} and \textit{actor} networks with weights $\theta^{Q}$ and $\theta^{\mu}$ \\
    Initialize the target networks as: $\theta^{\hat{Q}} \leftarrow \theta^{Q}, \theta^{\hat{\mu}} \leftarrow \theta^{\mu}$ \\
    \For{episode = 1, MaxEpisode}{
            $\triangleright$ Initialize \textbf{UAM Agent Environments}\\
            \For{time step = 1, T}{
            $\triangleright$ With probability $\pi(\mathcal{S}|\theta^{\mu})$ select a set of actions $\mathcal{A}$ for each $s^{m}_{t} \in \mathcal{S}$\\ 
            $\triangleright$ Execute actions at in \textbf{Simulation Environments} and observe reward $\mathcal{R}_{total}$ and the next set of states $\mathcal{S}'$\\
            $\triangleright$ Store the transition pairs $\xi = (\mathcal{S}, \mathcal{A}, \mathcal{R}_{total}, \mathcal{S}')$ in replay buffer $\Phi$\\
            \textbf{If} \textit{time step} \textbf{is update period and}  \textit{replay buffer} \textbf{is full enough to train, do followings:}\\
            $\triangleright$ Sample a mini-batch randomly from $\Phi$\\
            $\triangleright$ Set $y_{i} = r_{i} + \gamma \hat{Q}(s_{i}^{'}, \pi(s_{i}^{'}|\theta^{\hat{\mu}})|\theta^{\hat{Q}})$ \\
            $\triangleright$ Update the $\theta^{Q}$ by stochastic gradient descent to the loss function of \textit{critic} network: $L(\theta^{Q}) = \frac{1}{\varphi}\sum_{i}{(y_{i} - Q(s_{i}, a_{i}|\theta^{Q}))}^{2}$ \\
            $\triangleright$ Update the $\theta^{\mu}$ by stochastic gradient ascent with respect to the gradient of objective function in \textit{actor} network: \\
            $\nabla_{\theta}J(\theta^{\mu}) \approx \mathbbm{E}[{Q(s, a|\theta^{Q})\nabla_{\theta}\pi(s|\theta^{\mu})}]$ \\
            $\triangleright$ Update parameters of \textit{Target Networks} $\theta^{\hat{Q}}$ and $\theta^{\hat{\mu}}$
            }
        }
    \caption{Coordination of autonomous multi-UAM agents for reliable air transportation service.}
    \label{alg:actor critic}
\end{algorithm}

Next, we describe the training procedure of UAM agents. All UAM agents train their policy with multi-agent actor-critic DRL framework~\cite{lowe2017multi}. There are two independent neural networks, which are \textit{actor} and \textit{critic} networks. The parameters of \textit{actor} and \textit{centralized critic} network are denoted as $\theta^\mu_m$, $\forall m \in [1, M]$ and $\theta^Q$, respectively. Note that $M$ UAM agents have their own dependent $actor$ networks and share one $critic$ network in our proposed MADRL framework. By doing this strategy, CTDE-based agents use a centralized neural network when training offline, then execute online after sharing the trained neural network in a decentralized manner. Below is the detailed of CTDE-based training procedure.
%\begin{enumerate}
    %\item 
    We initialize the parameters of \textit{actor} network $\theta^\mu_m$, and \textit{critic} network $\theta^Q$ (line 1). Also, initialize parameters of target \textit{actor} network $\theta^{\hat \mu}_m$ and \textit{critic} network $\theta^{\hat Q}$ (line 2). The target network helps the learning process more stable by making the target value $y_i$ independent of the parameters.
    %\item 
    All UAM agents learn the policy by repeating the following procedure until all episodes are finished:
    (i) For every time step, all UAM agents take action by the policy of their own \textit{actor} networks $\pi_m(a_m|o_m;\theta^\mu_m)$. By taking action, they move to the next time step and make transition pairs $\xi = ({S, A, R_{total}, S^{'}})$, where the state is denoted by $S = \{o_m\}^M_{m=1}$. Sampled transition pairs are stored in experience replay buffer $\Phi$~\cite{mnih2013playing} (lines 4–8).
    (ii) After $\Phi$ is full enough to sample, UAM agents randomly sample $I$ transition pairs from $\Phi$ where $I$ is the size of the mini-batch. By training the neural network after the buffer is full enough to train, we can prevent biased training of the UAM agent to initial data. Experience replay buffer helps improve the learning performance of the UAM agent by reducing the continuity of the data used for training.
    (iii) Using $i$-th sampled transition pair in the mini-batch, the loss function $L(\cdot)$ is calculated with target value $y_{i}$ and action-value function $Q(s_{i}, a_{i};\theta^{Q})$ by Bellman optimality equation (line 9).
    (iv) Update the \textit{critic} network via gradient descent to reduce the value of the loss function (line 10-11). After that, update the \textit{actor} network $\theta^{\mu}_m$ via gradient ascent to increase the value of objective function $J(\theta^{\mu})$ (line 12-13).
    (v) Update the target parameters $\theta^{\hat{\mu}_m}$ and $\theta^{\hat{Q}}$ at specific periods (line 14).
%\end{enumerate}

\section{Performance Evaluation}
\subsection{System Settings}
In our environment, we considers 3D grid map which is $10\text{km}\times10\text{km}\times600\text{m}$. There are five different vertiports at (0km, 6km), (-6km, 1km), (6km, 1km), (-4km, -6km), (4km, -6km). At the beginning of the episode, all UAM agents are centered on the grid. Each episode has a total of 100 time steps, which is 100 minutes in real time. $N$ passengers randomly request air transportation services to any UAM agents in one of five vertiports. The service requested by the passenger is delivered to the UAM agent within a 3km communication cell radius of the vertiport~\cite{kwon2020multiagent}. If there is a seat in the UAM agent, it accepts the request and establishes a radio link. This paper considers CommNet-based and DNN-based neural artificial network consisted of $L$ dense layers. We also use the $\epsilon$-greedy method to explore the UAM agent to experience different actions in various states. Detailed descriptions of hyperparameters to configure environment are summarized in Table\,\ref{tab:setup}.

%To validate whether the proposed method cooperatively achieves the common goal of UAM agents, we check the performance of the proposed and the comparison method as follows.

The basic operations of our proposed method and the comparison method are described as follows.
\begin{itemize}
    \item \textit{Proposed Method:} All agent UAMs try to serve optimal air transportation service to passengers based on CommNet. They simultaneously communicate with each other to share their own rich experiences (i.e., observation information). Each UAM agent calculates its communication variables by averaging hidden variables (i.e., experiences) of other UAM agents.
    \item \textit{Comparison Method:} There are no inter-communication with UAM agents. In other words, all UAM agents try to provide air transportation service to users based on DNN. Thus, each UAM agent takes an action as if it were in a single-agent reinforcement learning environment. This paper uses independent Q-learning~\cite{tan1993multi} to train DNN-based comparison method.
\end{itemize}

\begin{table}[t!]
\scriptsize
\centering
\caption{Environmental hyper-parameters}
\renewcommand{\arraystretch}{1.0}
\begin{tabular}{l||r}
\toprule[1pt]
\textbf{Notation} & \textbf{Value} \\ \midrule
Number of UAM agents, $M$ & 4 \\
Number of passengers, $N$ & 25 \\
Dense layer, $L$ & 6 \\
Size of mini-batch, $I$ & 256 \\
Size of experience replay buffer, $\Phi$ & 10k \\
Discount factor, $\gamma$ & 0.99 \\
Initial value of epsilon, $\epsilon$ & 0.3 \\
The number of nodes in all layers & 64 \\
Annealing epsilon & 0.0001 \\
Learning rate & 0.0005 \\
Training epochs & 100k \\
Activation function & ReLU \\
Optimizer & Adam \\
\bottomrule[1pt]
\end{tabular}
\label{tab:setup}
\end{table}

\subsection{Evaluation Results}
%To evaluate the performance of the proposed method, we compare the performance with the comparison method in various aspects, i.e., (i) reward convergence, (ii) quality of air transportation service, and (iii) trained trajectories of each UAM agent in all methods.
\subsubsection{Reward Convergence}
\begin{figure}[t!]
    \centering
    \subfigure[Total.]{
    \includegraphics[width=0.65\linewidth]{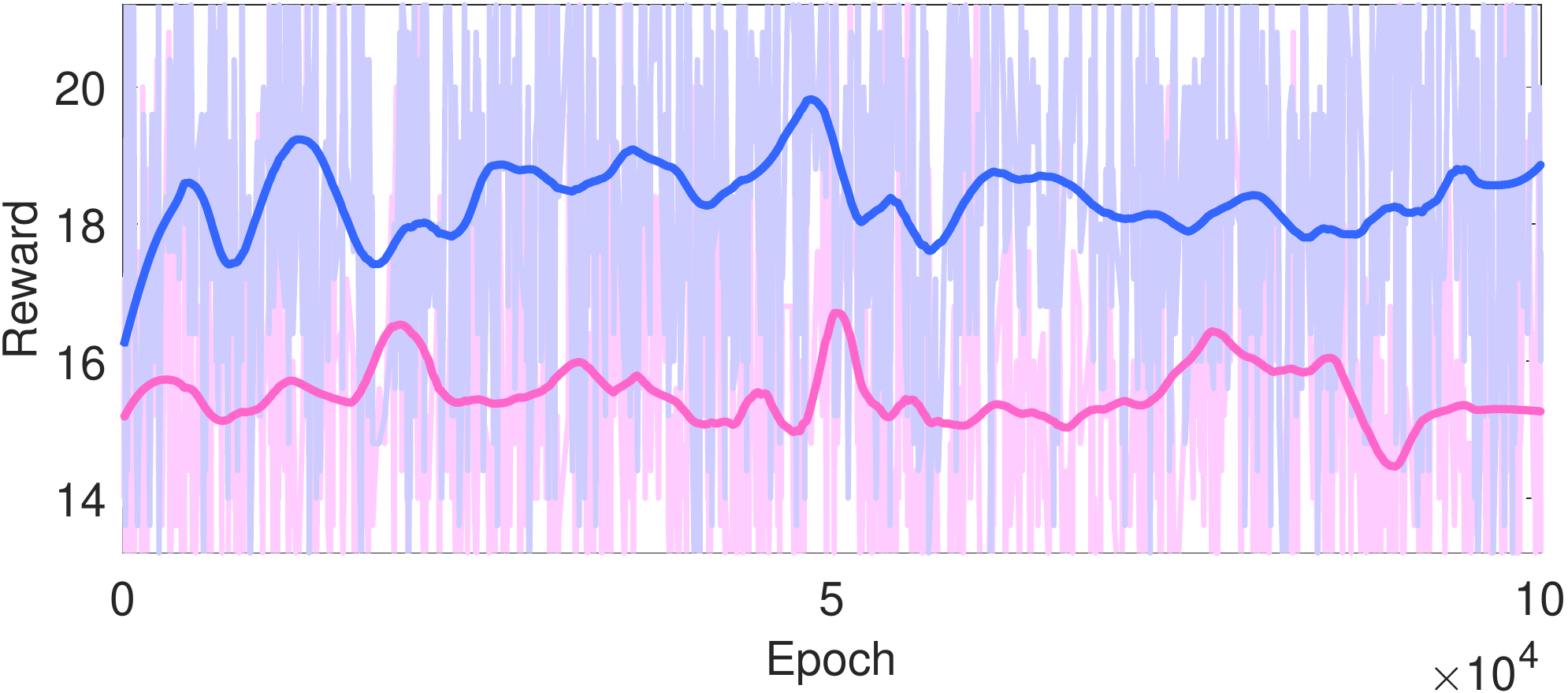}
    \label{fig:Total reward}
    }
    \subfigure[Average.]{
    \includegraphics[width=0.25\linewidth]{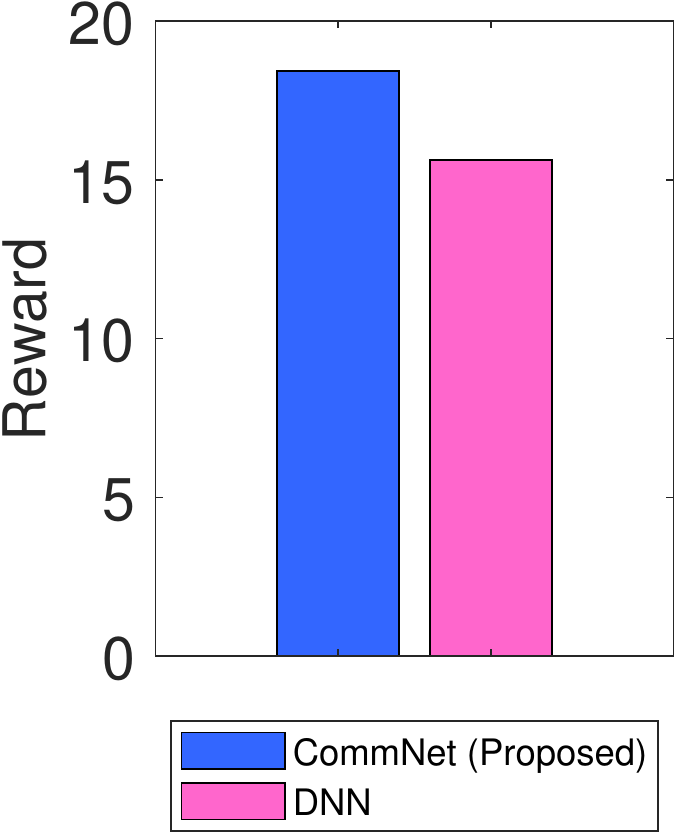}
    \label{fig:Average reward}
    }
    \caption{Fig.\,\ref{fig:Total reward} is the total reward at every training epochs and Fig.\,\ref{fig:Average reward} is the average reward in all training epochs.}
    \label{fig:rewards}
%    \vspace{-5mm}
\end{figure}

Fig.~\ref{fig:rewards} shows the tendency of reward convergence in all training epochs. In Fig.~\ref{fig:Total reward}, the proposed method shows a fast reward increase rate up to about the value 19 at the beginning of the training. The total reward in the proposed method finally converges at around [18,19]. Conversely, the comparison method does not show a conspicuous reward increase, and the total reward converges at around [15,16]. Furthermore, the average reward of the proposed method in Fig.~\ref{fig:Average reward} is about 18~\% higher than the comparison method. By virtue of the above results, our proposed method has more outstanding reward convergence performance than the comparison method thanks to inter-agent communications by CommNet.

\subsubsection{Quality of Air Transportation Service}
\begin{figure}[t!]
    \centering
    \subfigure[CommNet.]{
    \includegraphics[width=0.25\linewidth]{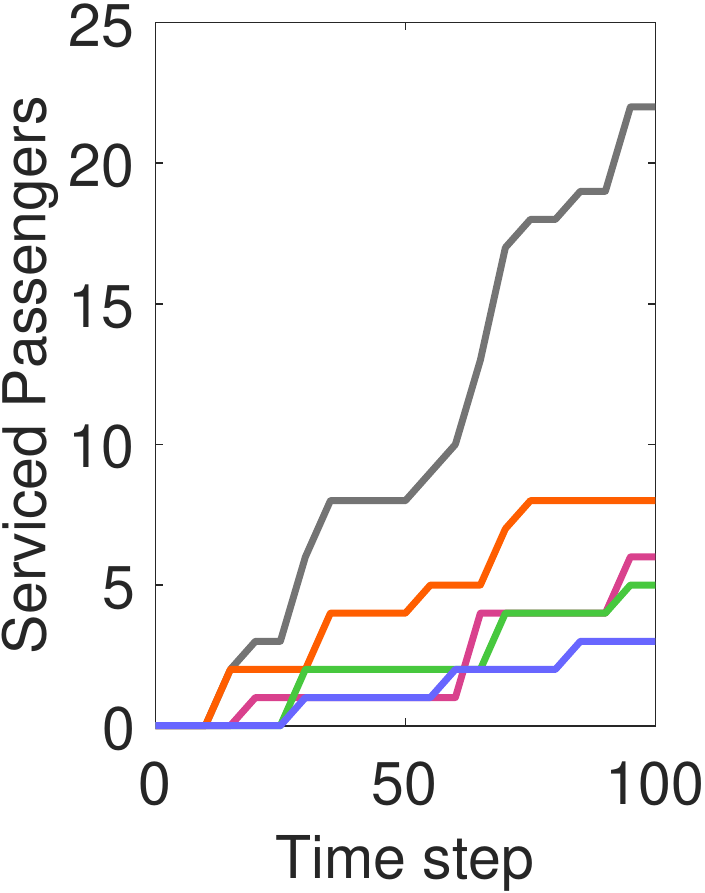}
    \label{fig:number of serviced users in Proposed}
    }
    \subfigure[DNN.]{
    \includegraphics[width=0.25\linewidth]{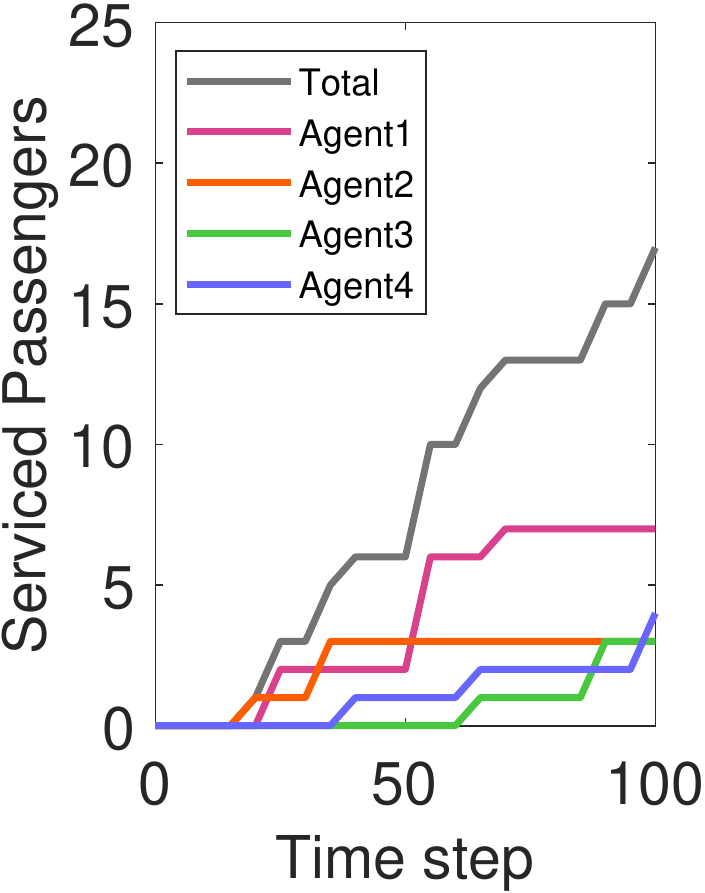}
    \label{fig:number of serviced users in Comp}
    }
    \subfigure[Waiting Time.]{
    \includegraphics[width=0.25\linewidth]{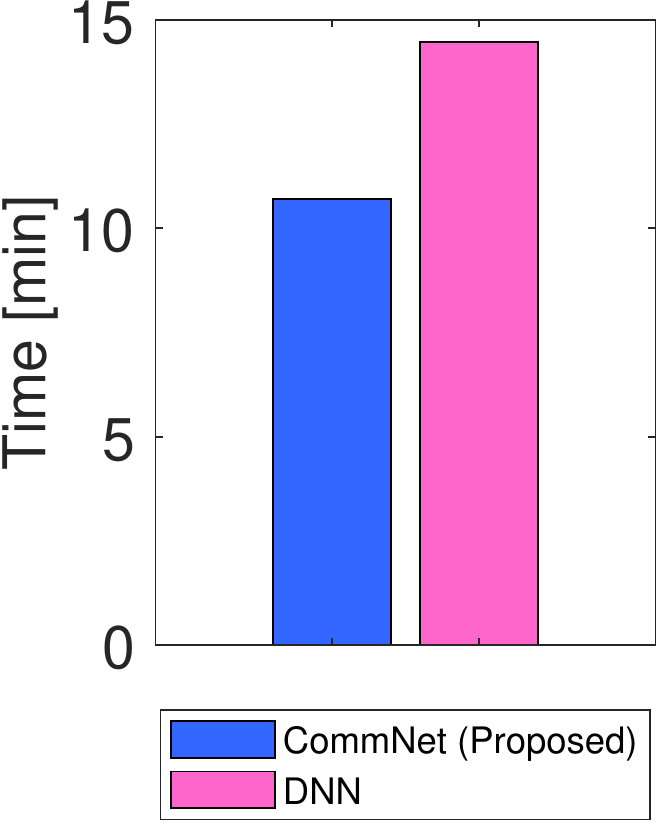}
    \label{fig:waiting time}
    }
    \caption{Fig.~\ref{fig:number of serviced users in Proposed}--\ref{fig:number of serviced users in Comp} are the number of serviced passengers by UAM agents in each method, and Fig.~\ref{fig:waiting time} is the waiting time per serviced passenger.}
    \label{fig:service quality}
\end{figure}

This paper considers the quality of air transportation service as the number of serviced passengers and waiting time per passenger. We will discuss about the service quality with Fig.~\ref{fig:service quality} as follows.
\begin{itemize}
    \item \textit{Number of Serviced Passengers:}
    Fig.~\ref{fig:number of serviced users in Proposed}$-$\ref{fig:number of serviced users in Comp} shows the number of passengers serviced by the UAM agent, i.e., arrive at their destination vertiport. Each UAM agent in the proposed method serves air transportation service to 6, 8, 5, and 3 passengers, and in the case of the comparison method, to 7, 3, 3, and 4 passengers. Finally, it is obvious that the proposed method serves approximately 30~\% more passengers than the comparison method.
    \item \textit{Waiting Time per Serviced Passenger:}
    Fig.~\ref{fig:waiting time} shows each method's overall average waiting time for serviced passengers. The passengers in the proposed method spend an average of 26~\% fewer times waiting to arrive at their destination vertiport than in the comparison method.
\end{itemize}

To put it succinctly, UAM agents in the proposed method provide higher quality air transport service to passengers than the comparison method in terms of the number of service passengers and overall waiting time.

\subsubsection{Trained Trajectories of Each UAM Agent}
\begin{figure}[t!]
    \centering
    \subfigure[Agent 1.]
    {
        \includegraphics[width=0.2\linewidth]{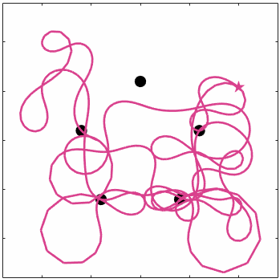}
        \label{fig:Proposed agent1}
    }
    \subfigure[Agent 2.]
    {
        \includegraphics[width=0.2\linewidth]{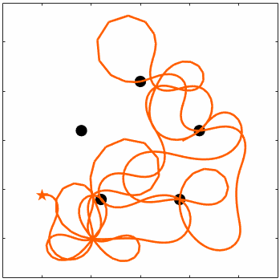}
        \label{fig:Proposed agent2}
    }
    \subfigure[Agent 3.]
    {
        \includegraphics[width=0.2\linewidth]{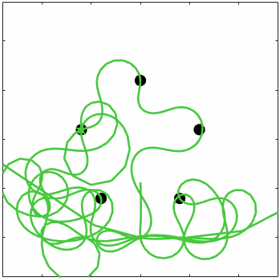}
        \label{fig:Proposed agent3}
    }
    \subfigure[Agent 4.]
    {
        \includegraphics[width=0.2\linewidth]{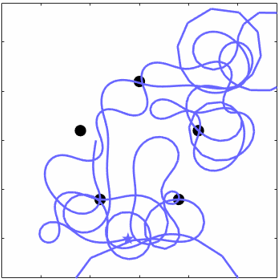}
        \label{fig:Proposed agent4}
    }
    \caption{The trajectories of CommNet-based UAM agents for all time step.}
    \label{fig:CommNet Trajectory}
\end{figure}

\begin{figure}[t!]
    \centering
    \subfigure[Agent 1.]
    {
        \includegraphics[width=0.2\linewidth]{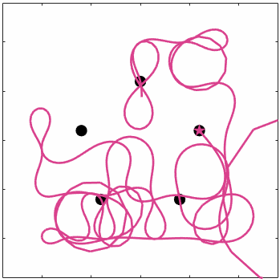}
        \label{fig:comp1 agent1}
    }
    \subfigure[Agent 2.]
    {
        \includegraphics[width=0.2\linewidth]{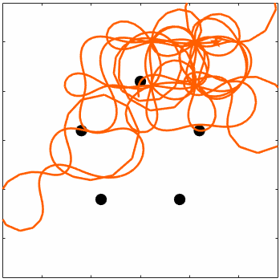}
        \label{fig:comp1 agent2}
    }
    \subfigure[Agent 3.]
    {
        \includegraphics[width=0.2\linewidth]{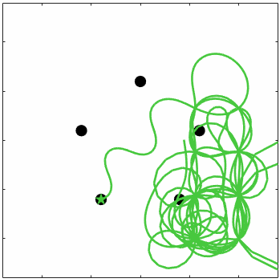}
        \label{fig:comp1 agent3}
    }
    \subfigure[Agent 4.]
    {
        \includegraphics[width=0.2\linewidth]{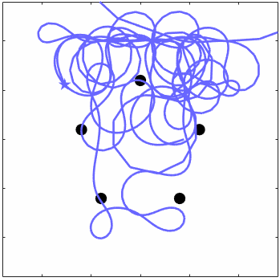}
        \label{fig:comp1 agent4}
    }
    \caption{The trajectories of DNN-based UAM agents for all time step.}
    \label{fig:DNN Trajectory}
\end{figure}

We describe the trajectories of UAM agents after training their policies. In Fig.~\ref{fig:CommNet Trajectory}--\ref{fig:DNN Trajectory}, the number of vertiports each UAM agent has ever landed on is 4, 4, 5, and 4, and in the case of the comparison method, 4, 3, 3, and 3, respectively. UAM agents of the proposed method arrived at an average of about 31~\% different vertiports than the comparison method. Moreover, all UAM agents of the proposed method cruise evenly across the entire map. However, UAM agents in the comparison method only cruise in a limited area except for $b_1$. Taking these results into account, CommNet helps UAM agents learn more about their environment. In other words, CommNet-based UAM agents efficiently achieve a common goal by communicating different information that UAM agents observe.
Our proposed trajectories those are plotted in Fig.~\ref{fig:CommNet Trajectory} is based on nonlinear control.
If we can consider linear trajectory control and optimization, it would be better in terms of moving distance minimization. However, it is not realistic in the view points of practical aircraft control. Therefore, it can be verified that our approach is more practical even though the moving trajectories have additional nonlinear control movement. 

\section{Conclusions}
This paper proposes a novel multi-UAM cooperative CTDE-based MADRL algorithm for reliable and efficient passenger delivery in UAM networks. To train the centralized \textit{critic} neural network, we define optimization standards based on the reward function in MDP, considering the actual specifications of the UAM model. According to the performance evaluation results in various ways, we verify that the proposed algorithm outperforms other existing algorithms in terms of the number of serviced passengers increase ($\approx 30$\%) as well as the waiting time per serviced passenger decrease ($\approx 26$\%). %The future work of this study is to consider the real map, i.e, the real position of vertiports, sudden obstacles.

\section*{Acknowledgements} 
This research was funded by National Research Foundation of Korea (2022R1A2C2004869, 2021R1A4A1030775). \textit{(Corresponding authors: Soyi Jung, Joongheon Kim)}.

% \cite{icufn22kim}
% \cite{ton201608kim}
% \cite{jsac201806choi}
% \cite{access202106park}
% \cite{tvt201905shin}
% \cite{tvt202106jung}
% \cite{tii202005shin}
% \cite{tii202210yun}
% \cite{icoin21jung}
% \cite{icte202103yun}
% \cite{jcn2022lee}

\bibliographystyle{IEEEtran}
\bibliography{ref_aimlab, ref_icc}

\end{document}